\documentclass{article}
\usepackage[utf8]{inputenc}
\usepackage{graphicx}
\usepackage{fullpage}
\usepackage{subfigure}

\newcommand{\eat}[1]{}

\newcommand{\aba}{AtlasBuilder actor}
\newcommand{\sa}{Sampler actor}
\newcommand{\wa}{Writer actor}
\newcommand{\rmc}{rigid molecular component}

\title{Parallel Exploration of Directed Acyclic Graphs using the Actor Model}
\author{Rahul Prabhu \and Amit Verma \and Meera Sitharam}

\begin{document}

\maketitle

\eat{
\section{Structure}
\begin{enumerate}
    \item Introduction - Parallel exploration of DAG, Application areas, Actor model (from grant), example easal
    \item What/Why Actor Framework - explain actor framework, differences from MPI and other frameworks, advantages in terms of simplicity of code due to lack of locking etc., ability to use distributed systems (rather than depending on a cluster, like in MPI), performance, I/O, etc.
    \item Architecture - Parallel exploration of DAG, use generic terms for the atlasbuilder and sampler, redraw the architecture diagram, give hints as to how this can be done in a distributed fashion
    \item Results - Experimental Setup -- details about EASAL? Hipergator, plots, challenges and optimizations (connect it to the larger context if possible)
    \item Discussion/Conclusion
\end{enumerate}
}

\begin{abstract}
In this paper we describe a generic scheme for the parallel exploration of directed acyclic graphs
starting from one or more `roots' of the graph. Our scheme is designed for graphs with the following
properties, (i) discovering neighbors at any node requires a non-trivial amount of computation, it is not
a simple lookup;  (ii) once a node is processed, all its neighbors are discovered; (iii) each node can be 
discovered through multiple paths, but should only be processed once. Several computational problems 
can be reduced to traversing such graphs, where the goal is to explore the graph and build 
a traversal roadmap. As a proof of concept for the effectiveness of our scheme at achieving speedup due to parallelism,
we implement the scheme for the  parallel exploration of assembly landscape using the EASAL methodology.
\end{abstract}

\section{Introduction}

In this paper we describe a generic scheme for the parallel exploration of directed acyclic graphs
starting from one or more `roots' of the graph. Our scheme is designed for graphs with the following
properties, (i) discovering neighbors at any node requires a non-trivial amount of computation, it is not
a simple lookup;  (ii) once a node is processed, all its neighbors are discovered; (iii) each node can be 
discovered through multiple paths, but should only be processed once. Several computational problems 
can be reduced to traversing such graphs, where the goal is to explore the graph and build 
a traversal roadmap. For example, in terrain mapping using robots, the mapped terrain is typically represented as 
a graph, where the various features of the landscape are the nodes of the graph and the path traversed 
by the robot to get between the features are the edges \cite{ota2006multi}. In the exploration of energy 
landscapes in chemistry, the macrostates are the nodes and the boundary relationships between the macrostates 
are the edges \cite{PrabhuEtAl2020}.

Our parallel scheme is based on the \emph{actor model}, which is a computational model for concurrent 
programming, in which the basic unit of computation is an abstract entity called an \emph{actor} 
(the actor model is discussed in more detail in Section \ref{sec:arch}). 
As a proof of concept for the effectiveness of our scheme at achieving speedup due to parallelism,
we implement the scheme for the  parallel exploration of assembly landscape using the EASAL methodology.
We use the C++ Actor Framework (CAF) \cite{chs-rapc-16, cshw-nassp-13} for the implementation.
The parallel implementation, running on University of Florida's Hipergator super computer, on 
Intel(R) Xeon (TM) Gold 6142 series processors from the Skylake family series of processors gives 
near linear speedup in the number of compute cores (as compared to the sequential implementation 
of the EASAL methodology \cite{easalSoftware}) and can roadmap several millions nodes in minutes, 
where the sequential version took hours.

\subsection{Related work: background on EASAL}

\eat{GPSA: A Graph Processing System with Actors
https://arxiv.org/vc/arxiv/papers/1008/1008.1459v8.pdf}

\label{sec:easal_background}
Molecular assembly configuration spaces are topologically complex. Even when modeling 
just two assembling molecules, the space is 5 dimensional, with it growing exponentially 
with the number of assembling molecular units. Not just that, they are topologically 
complex with a lot of disconnected components and non-convexity. Traditional methods for 
analyzing these energy landscapes - such as Monte Carlo sampling and Molecular Dynamics 
simulations - are non-ergodic, not deterministic, and computationally intensive.

The EASAL methodology \cite{PrabhuEtAl2020} analyzes landscapes of assembly driven 
by short-range pair-potential interactions such as Lennard-Jones potentials. 
This method is resource light, and generates the entire potential energy landscape 
or configuration space of assembly driven by short 
range potentials. EASAL takes as input $k$ rigid molecular components with at most $n$ 
atoms each. Pairwise distance interval parameters, which have been geometrized and 
discretized. It can optionally take global constraints on 
the overall configuration which mimic an implicit solvent. 

To atlas assembly configuration spaces, EASAL employs 3 main strategies 
The first strategy it employs is partitioning the assembly landscape into nearly-equipotential 
energy regions regions called \emph{active constraint regions} or \emph{macrostates}. Each region is 
characterized by configurations that have a particular set of ‘active’ constraints (atom pairs, 
one from each molecule in the LJ well). After partitioning them, we organize these regions in 
the form of a directed acyclic graph (DAG). We put a directed arrow from a parent region to a 
child region with one additional active constraint. This gives us a topological \emph{roadmap} 
or an atlas of the configuration space. We label each active constraint region with an 
\emph{active constraint graph}, the vertices are atoms in the two molecules, and edges are of two 
types, ones within the same molecule and the ones going across. We use combinatorial rigidity to 
analyze this graph and determine the effective dimension of the region. Consequently, the effective 
dimension becomes a proxy for the energy level of the macrostate. Thus, in addition to organization 
as a DAG, we are able to stratify the configuration space. So, this amorphous ugly looking landscape 
now has some structure to it, which makes it somewhat easier to handle.

We are interested in finding the 0-dimensional configurations that form the bottoms of potential energy 
basins. As its second strategy, EASAL, tarting from the interior of a higher dimensional region, 
recursively searches for regions of exactly one dimension less. This has a higher chance of success than 
trying to find the lowest energy regions directly.

The third strategy is the use of  a novel distance-based parameterization called Cayley parameterization. 
Cayley parameters many-to-one map a d-dimensional active constraint region to a convex d-dimensional 
region called the Cayley space. This makes sampling active constraint regions easier. In addition, 
computing the inverse map from Cayley configurations to Cartesian configurations is as easy as solving 
quadratic equations. Each realization corresponds to a chirotrope or flip.

Taking advantage of EASAL’s first strategy, i.e., partitioning of the assembly landscape into active constraint 
regions, and leveraging the fact that these regions can be explored independently of each other, we develop
our parallel scheme for roadmapping assembly landscapes.

\eat{
In this paper we describe a parallel roadmapping scheme for the exploration
of assembly landscapes. Taking advantage of EASAL's partitioning of the
assembly landscape into active constraint regions, and the fact that these
regions can be explored independently of each other, we use a modified version
of the core EASAL algorithm \cite{Ozkan:toms} to roadmap assembly landscapes in
parallel, while avoiding repeat processing active constraint regions.

While, the implementation described in this chapter is for the exploration of
assembly landscapes, the parallel scheme described is more generally
applicable. In fact, it can be used for mapping out any Directed Acyclic
Graph (DAG) that has the following properties:

Several naturally occurring problems can be modeled as a DAG exploration
problem. Examples include (i) roadmapping in robotics and configuration space
analysis, (ii) simulating animal/insect behavior (ant colony optimization
algorithms), and (iii) parallel exploration of a region by multiple robots. In
this chapter we describe the parallel algorithm as applied to the roadmapping of
the configuration space of assembling molecules.}

\noindent \textbf{Organization:} The rest of the paper is organized as
follows. Section \ref{sec:arch} describes the architecture of the parallel
roadmapping scheme. Section \ref{sec:results} gives the results for the
roadmapping of assembly landscapes with the molecules described in the preivous
chapter as input. Section \ref{sec:discussion} discusses the design
decisions taken for the EASAL implementation of our methodology.

\eat{
\section{Related Work}
\label{sec:RelatedWork}
Graph Theory Perspective - Parallel exploration of graphs
1. A work efficient algorithm for parallel unordered depth-first search
2. Parallel Depth-first search for directed acyclic graphs

Applications perspective - 
Multi-Agent Path finding 
1. A polynomial time algorithm for non-optimal multi-agent path finding
2. Global planning on th emars expolration rovers
3. Using quadtrees for realtime pathfinding in indoor environments
4. Heirarchical path planning for mulit-size agents in heterogeneous environments

\section{Why Actor Model}
\label{sec:whyactor}
\subsection{Actor Model}
\label{sec:actorframework}

}

\section{Architecture}
\label{sec:arch}
This section discusses the architecture of our parallel roadmapping scheme 
in the EASAL methodology using the actor model.
The actor model is a concurrent programming model, in which the basic unit of
computation is an abstract entity called an \emph{actor}. Actors communicate
asynchronously through message passing and have well defined behaviors for each
message they receive. In response to messages, actors can make local decisions,
reply to messages, or spawn more actors. Actors don't share state, avoiding
critical sections and race conditions commonly found in concurrent programs.
This obviates the need for synchronization through locks, leading to less
complicated code and better performance. 

Our architecture uses three different types of actors, an atlasbuilder actor 
to keep track of the atlas and ensure no repeated sampling of active constraint 
regions, several sampler actors for sampling the active constraint regions parallely, 
and a writer actor to write the output to disk. Each of these actors are explained 
in detail below.

\subsection{Actors}
\label{sec:actors}
Our architecture uses three different types of actors, an atlasbuilder actor to
keep track of the atlas and ensure no repeated sampling of active constraint
regions, several sampler actors for sampling the active constraint regions
parallely, and a writer actor to write the output to disk. Each of these actors
are explained in detail below.

\subsubsection{AtlasBuilder Actor}
\label{sec:ABActor}
The \aba\ is the central actor that starts the roadmap exploration.
It also keeps the only copy of the entire roadmap, which it uses to avoid
repeat exploration of active constraint regions.
The \aba\ has the following data structures in its local state.

\begin{enumerate}
\item \textbf{Atlas}: The atlas data structure the a central singleton data
structure which stores the atlas.

\item \textbf{New regions queue}: A queue of newly discovered regions that need
to be sent to samplers for sampling and exploration. Depending on the sampling
policy, either DFS or BFS, this queue is either a multi-level queue (the levels
are the dimensions of the regions) or a single-level queue.

\item \textbf{Root graphs list}: The list of all rootgraphs to be sampled.

\item \textbf{Writer actor}: Pointer to the singleton \wa.
\end{enumerate}

The \aba\ accepts to the following messages. 

\begin{enumerate}
\item \textbf{Start:} Upon receiving the \emph{Start} message, the \aba\
initializes the atlas data structure, the new regions queue, and
initializes book keeping counters. Next, it creates the initial root 
graphs and queues them for sampling and exploration. Finally, it sends itself 
the Ready message to start the roadmapping process.

\item \textbf{Ready:} Upon receiving the \emph{Ready} message, the \aba\
adds all the root graphs in the root graphs list to the new regions queue and 
sends a Process message to itself.

\item \textbf{Process:} Upon receiving the \emph{Process} message, the \aba\
iterates through the new regions queue, spawns a new \sa\ and sends a
\emph{Sample} message to it with the element on at the front of the queue as
input. The \aba\ keeps doing this till either the new regions queue become
empty or the number of active samplers hasn't gone beyond the maximum \sa\
allowed.

\item \textbf{Sampling Result:} The \emph{Sampling Result} is sent to the \aba\
by the \sa\ when it discovers a new boundary point and thus potentially a new
active constraint region. Upon receiving the Sampling Result message, the \aba\
checks the atlas data structure to determine if the active constraint region in
the message is newly found or if it had been previously discovered. If the
active constraint region it is newly found, it adds it to the new regions
queue. If the region was previously found, it simply sends a write witness
message to the \wa with it. In either case, it computes the boundary region
adjacent to the boundary point and returns it to the \sa.

\item \textbf{Done:} The \emph{Done} message is sent to the \aba by the \sa\
when it completes sampling the acitve constraint region given to it
as input. Upon receiving the Done message, the \aba\ decrements its counter 
of the number of active samplers, sends itself a process message, and responds with a Kill message.
\end{enumerate}

\subsubsection{Sampler Actor}
\label{sec:samplerActor}
\sa s sample a single active constraint region given to them as input. In the
process of sampling, they relay any newly found active constraint region back
to the \aba\ and send the sampled data to be written to the \wa. At any given
time, there are several (up to the maximum number given as input by the user)
active \sa s.

A \sa\ has the following data structures in its local state.
\begin{enumerate}
\item \textbf{Active constraint region}: The active constraint region sent
as input from the \aba\ that needs to be sampled. All newly discovered
configurations, as a result of sampling, are stored in this data structure.
\item \textbf{Writer actor}: Pointer to the singleton \wa.
\end{enumerate}

The \sa\ responds to the following messages
\begin{enumerate}
\item \textbf{Sample:} Upon receiving this message, the \sa\ starts sampling
an active constraint region received as input. During sampling, any new regions found
are sent to the \aba\ to be stored in the atlas and for boundary information
retrieval.

\item \textbf{Boundary Result:} Upon receiving this message, the \sa\ updates 
its active constraint region data structure to indicate appropriate 
atlas node numbers for boundary Cayley points. 

\item \textbf{Kill:} Upon receiving this message, the \sa\ exits.
\end{enumerate}
\subsubsection{Writer Actor}
\label{sec:writerActor}

The \wa\ is a singleton actor that accepts the atlas data structure from the \aba\ and 
active constraint region data structures from \sa s and writes them to the disk. 

The \sa\ responds to the following messages
\begin{enumerate}
\item \textbf{Write Sample Points:} Upon receiving this message, the \wa\ opens
the appropriate node file and writes the entire active constraint
region to memory.
\end{enumerate}

\subsection{Workflow}
\label{sec:workflow}
\begin{figure}[htpb]
\centering
\includegraphics[scale=0.5]{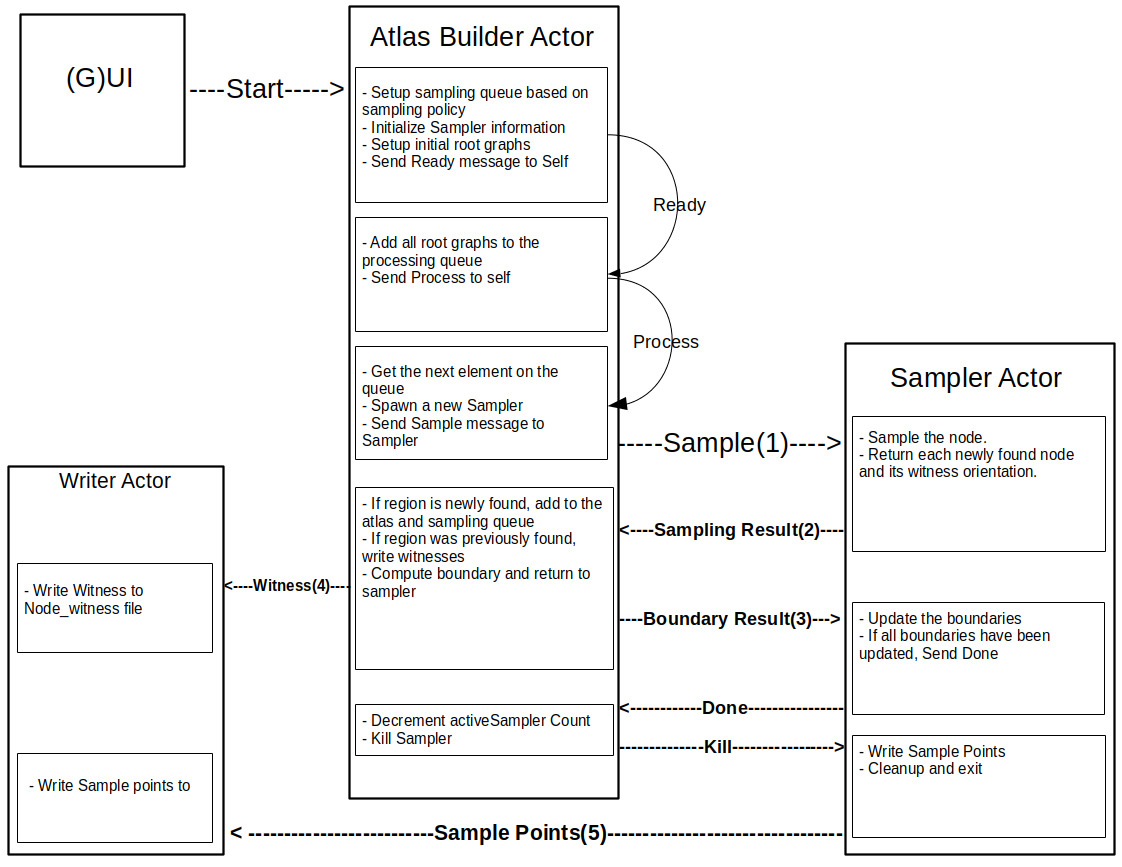}
\caption[Workflow of the parallel EASAL algorithm]{Workflow of the parallel EASAL algorithm}
\label{fig:workflow}
\end{figure}

Figure \ref{fig:workflow} shows the overall workflow of the parallel EASAL
algorithm. The algorithm starts with the main function sending a Start message
to the \aba. Upon receiving the Start message, the \aba\ initializes the atlas
data structure, the new regions queue, and initializes book keeping counters.
Next, it creates the initial root graphs and queues them for sampling and
exploration. Finally, it sends itself the Ready message to start the
roadmapping process. Upon receiving the \emph{Ready} message, the \aba\ adds
all the root graphs in the root graphs list to the new regions queue and sends
a Process message to itself. Upon receiving the \emph{Process} message, the
\aba\ iterates through the new regions queue, spawns a new \sa\ and sends a
\emph{Sample} message to it with the element on at the front of the queue as
input. The \aba\ keeps doing this till either the new regions queue become
empty or the number of active samplers hasn't gone beyond the maximum \sa\
allowed.

A \sa\ starts when it receives an active constraint region, identified by its
active constraint graph as input from the \aba. During sampling, any new
regions found are sent to the \aba\ to be stored in the atlas and for boundary
information retrieval. Upon receiving this message, the \sa\ updates its active
constraint region data structure to indicate appropriate atlas node numbers for
boundary Cayley points. 

Upon receiving the Sampling Result message, the \aba\
checks the atlas data structure to determine if the active constraint region in
the message is newly found or if it had been previously discovered. If the
active constraint region it is newly found, it adds it to the new regions
queue. If the region was previously found, it simply sends a write witness
message to the \wa\ with it. In either case, it computes the boundary region
adjacent to the boundary point and returns it to the \sa.

Once sampling is finished, the \sa\ sends all sampled points to the \wa\ to be
written to the disk and sends a Done message to the \aba. In response to the
Done message, the \aba\ decrements its counter of the number of active
samplers, sends itself a process message, and responds with a Kill message.
Upon reciept of the Kill message from the \aba, in repsonse to the Done
message, the \sa\ exits.

\section{Results}
\label{sec:results}

\begin{figure}[htpb] \centering
\subfigure[]{\label{fig:6Straight}\includegraphics[scale=0.08]{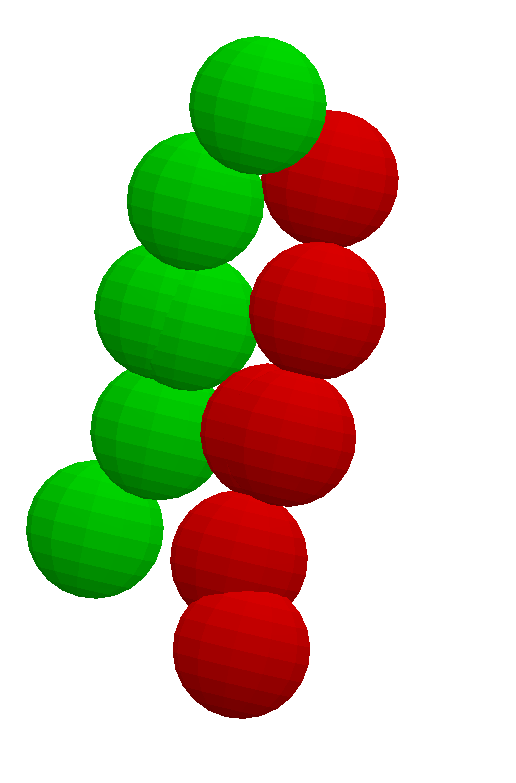}}
\subfigure[]{\label{fig:6Pocketed}\includegraphics[scale=0.08]{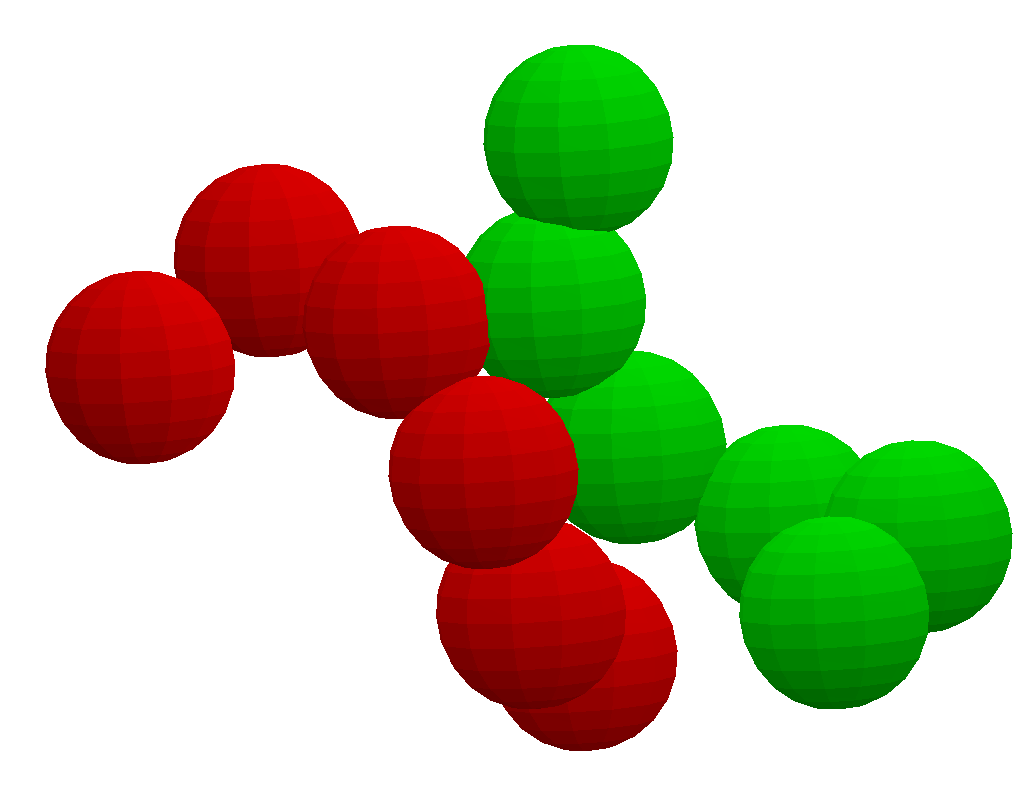}}
\subfigure[]{\label{fig:10Pocketed}\includegraphics[scale=0.08]{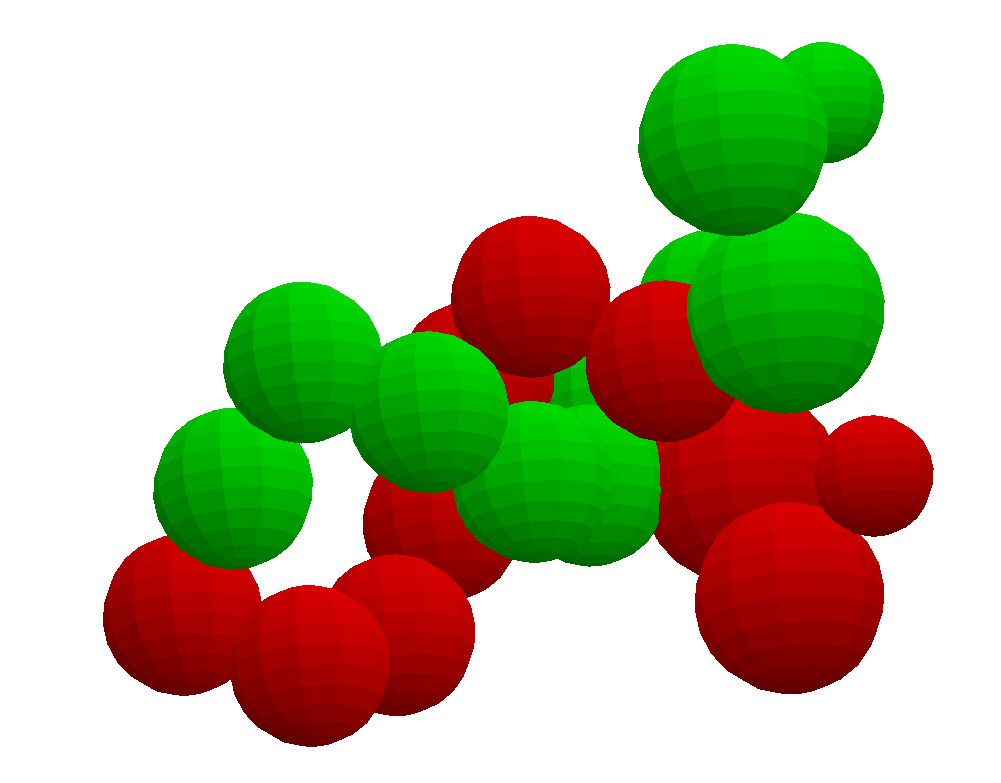}}
\subfigure[]{\label{fig:20Pocketed}\includegraphics[scale=0.08]{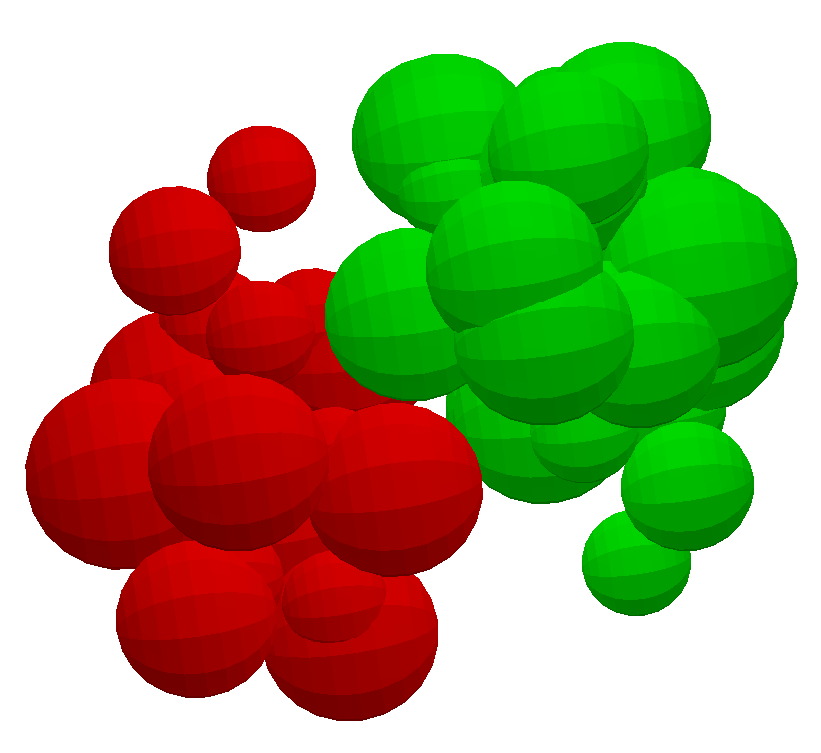}}
\subfigure[]{\label{fig:20Straight}\includegraphics[scale=0.08]{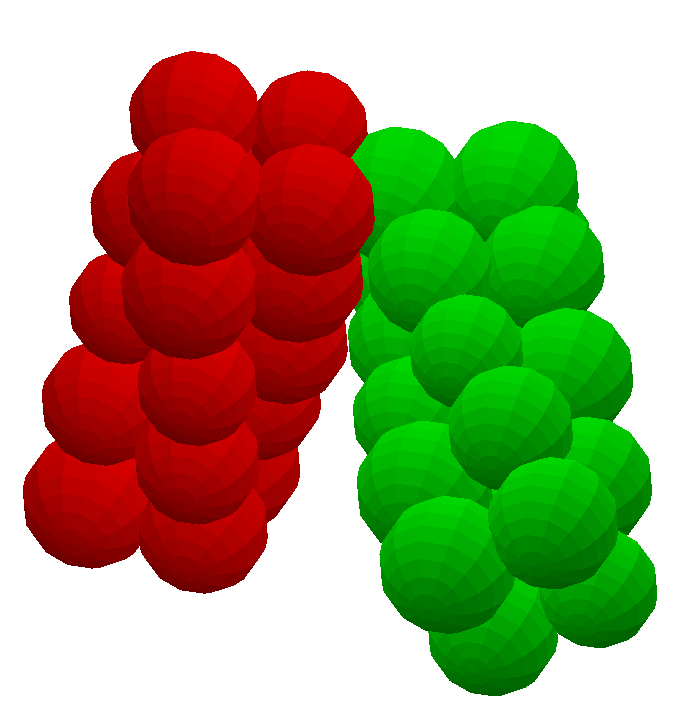}}
\subfigure[]{\label{fig:42Pocketed}\includegraphics[scale=0.08]{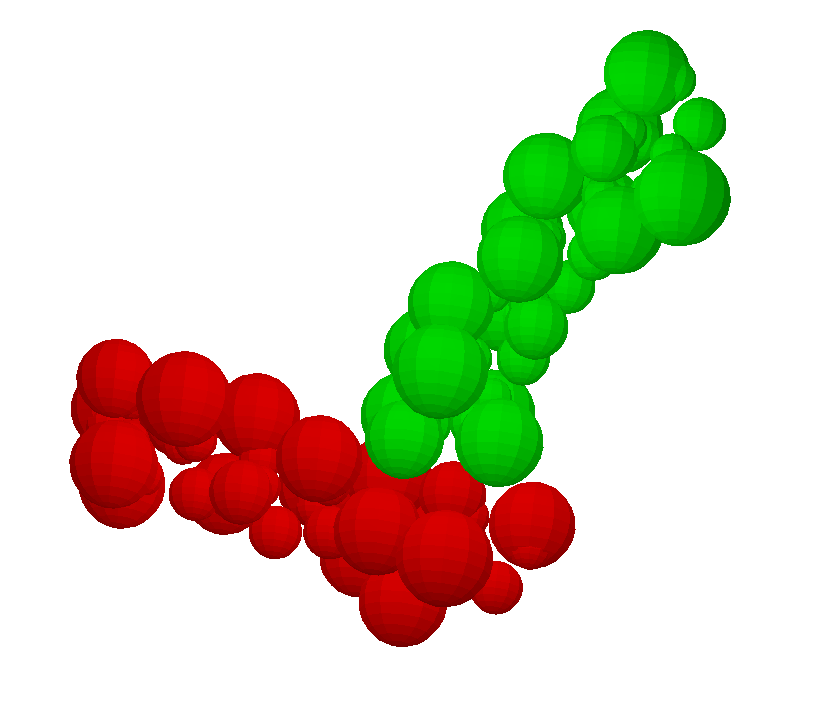}}
\caption{\scriptsize \textbf{List of \rmc s used in the experiments, illustrating different input 
shape variables}: (a) Narrow Convex (6 Atoms). (b) Narrow Concave (6
Atoms). (c) Narrow Concave (10 Atoms) (d) Wide Convex (20 Atoms). (e) Wide
		Concave (20 Atoms). (f) Wide Concave (42 Atoms) (see text in Section \ref{sec:results:expSetup}).}
\label{fig:inputMolecules}
\end{figure}

Results demonstrating the parallel EASAL algorithm's capabilities and high
computational efficiency have been generated by the opensource software
implementation EASAL (Ecient Atlasing and Search of Assembly Landscapes, see
software \cite{easalSoftware}, video \cite{easalVideo} and user guide
\cite{easalUserGuide}). The experiments were performed on assembly systems 
described in Section \ref{sec:results:expSetup}, each with $k = 2$ input 
\rmc s, set up to cover a wide variety of geometric shapes (input
shape variables, including number of atoms, concavity and width, see Figure
\ref{fig:inputMolecules}).

\subsection{Experimental Setup}
\label{sec:results:expSetup}
\begin{figure}[htpb]
\centering
\includegraphics[scale=0.4]{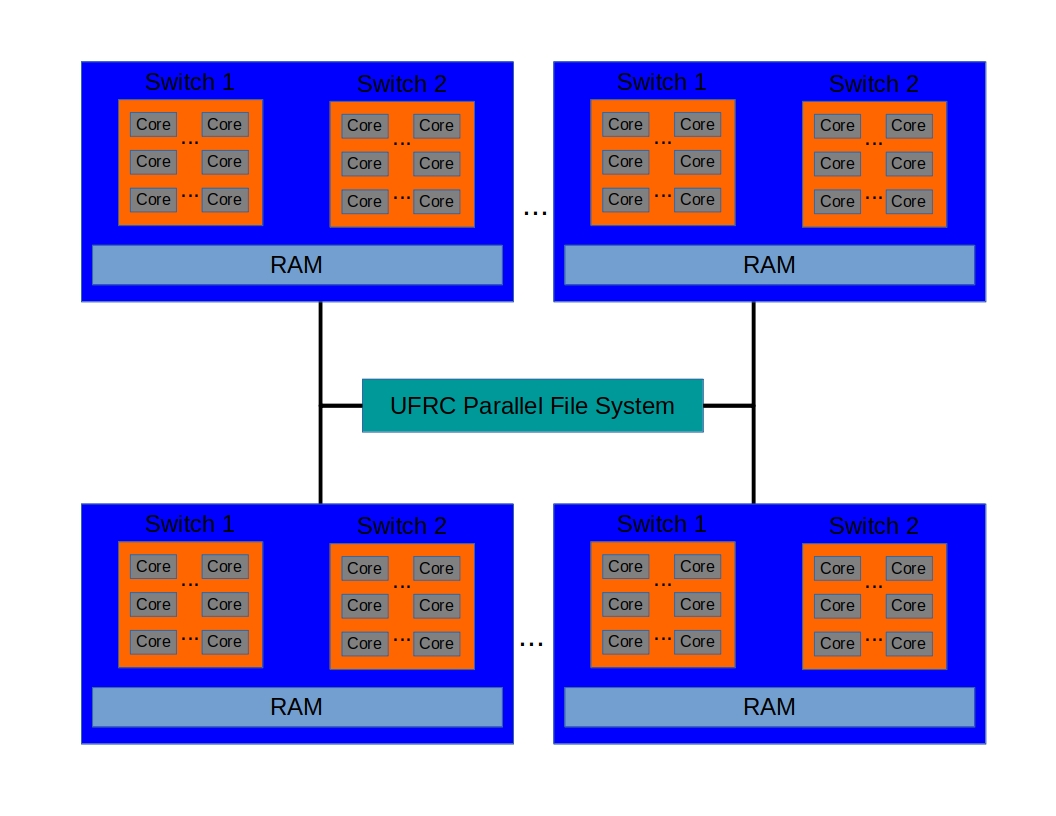}
\caption[Schematic diagram of the hipergator cluster]{Schematic diagram of the
hipergator cluster on which the experiments were performed. All the runs were
performed on cores belonging to a single switch in a single node of the
cluster. See Section \ref{sec:results:expSetup}.}
\label{fig:cluster}
\end{figure}

The experiments were run on the Hipergator supercomputer with a varying number
of Intel(R) Skylate(TM) series processors all on the same switch, in a single
node of the cluster, to reduce latency in mesage passing (see Figure
\ref{fig:cluster}. Suitable g++ compiler optimizations were used on the code to
enhance its performance.

The primary computation performed was atlas generation. For the input molecules
described earlier, we generate the complete atlas. To be able to compare the
results across \rmc s, we perform normalizations similar to what was done in
the single threaded experiments in the paper \cite{PrabhuEtAl2020}. In particular,
we fix the ratio of the average atom radius to the sampling step size $t$. We
additionally fix the width of the Lennard-Jones' well, $\overline{\delta_{ab}}
- \delta{ab}$ for an atom pair $(a, b)$,
with radii $\rho_a$ and $\rho_b$, to $0.25 * (\rho_a + \rho_b)$. Each of
these input assembly systems was sampled with 3 different values of $t$, to
analyze the effects of step size on the sampling time. In addition, each input
was run 10 times and the results were averaged to account for variabilities in
Hipergator loads.

\eat{
\subsection{Key Measurements}
\label{sec:results:keyMeasurements}
We use the following key measurements to test the capabilities of parallel
EASAL.

\begin{enumerate}
\item {\bf Total Sampling Time:} is the total wall time taken to sample
the entire atlas.
\item {\bf Rate of Macrostate discovery:} Is the number of macrostates 
discovered per second.
\end{enumerate}
}

\subsection{Verifying the Speedup of the Parallel Implementation}
\label{sec:results:verifyingSpeedup}

\begin{figure}[htpb]
\centering
\includegraphics[scale=0.1]{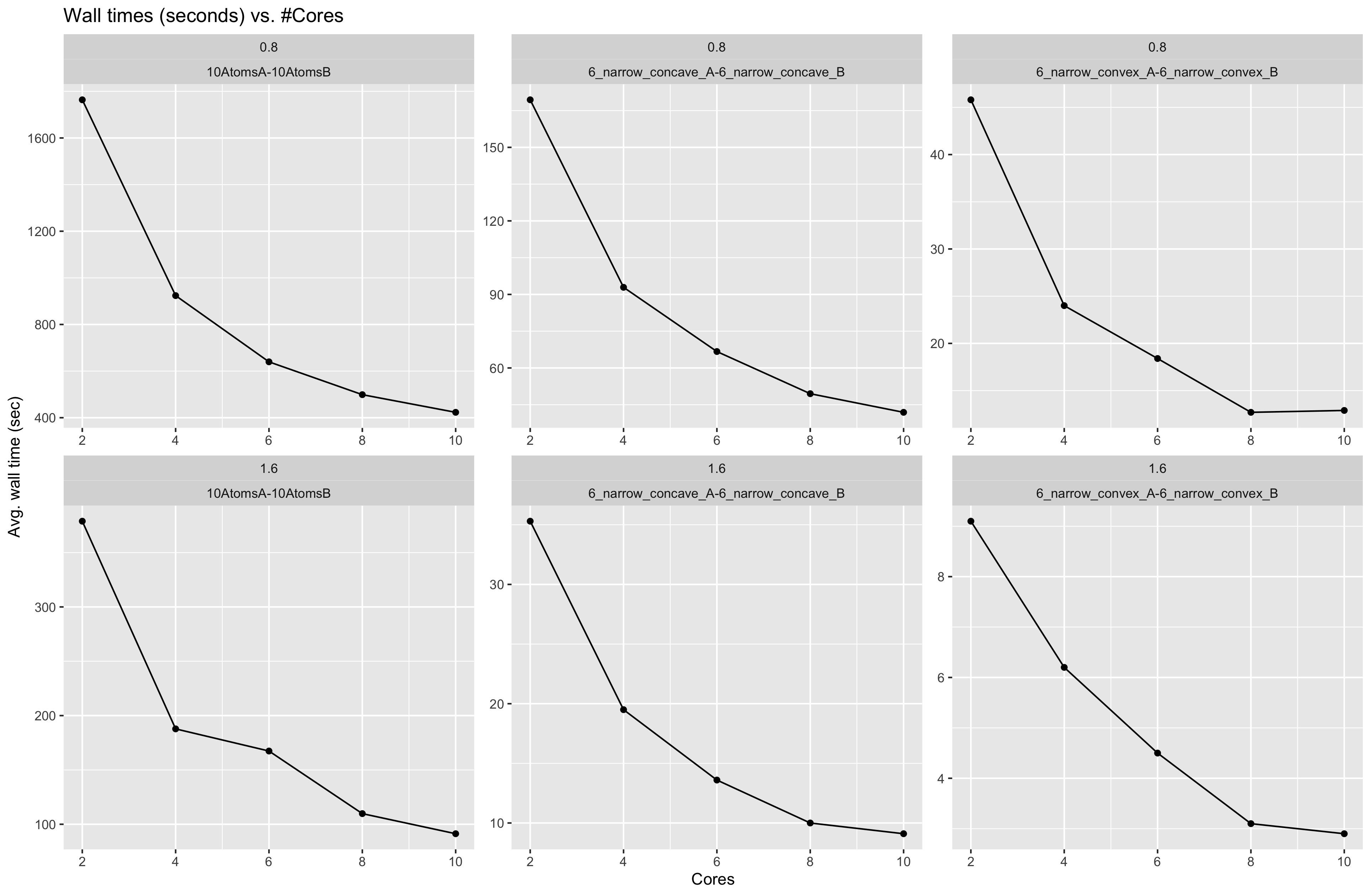}
\caption[Compute cores vs Atlasing time]{Shows the number of compute cores on the x-axis
and the average time required for sampling the entire atlas of the two
molecule mentioned at the top of every sub-figure, for two different step
sampling sizes. 
}
\label{fig:avgWallTime}
\end{figure}
Figure \ref{fig:avgWallTime} pltos the number of compute cores on the x-axis
and the average time required for sampling the entire atlas of the two
molecule mentioned at the top of every sub-figure, for two different step
sampling sizes. As can be seen from the figure, the atlasing time
reduces dramatically with increase in the number of cores and shows signs
of plateauing.

\begin{figure}[htpb]
\centering
\includegraphics[scale=0.1]{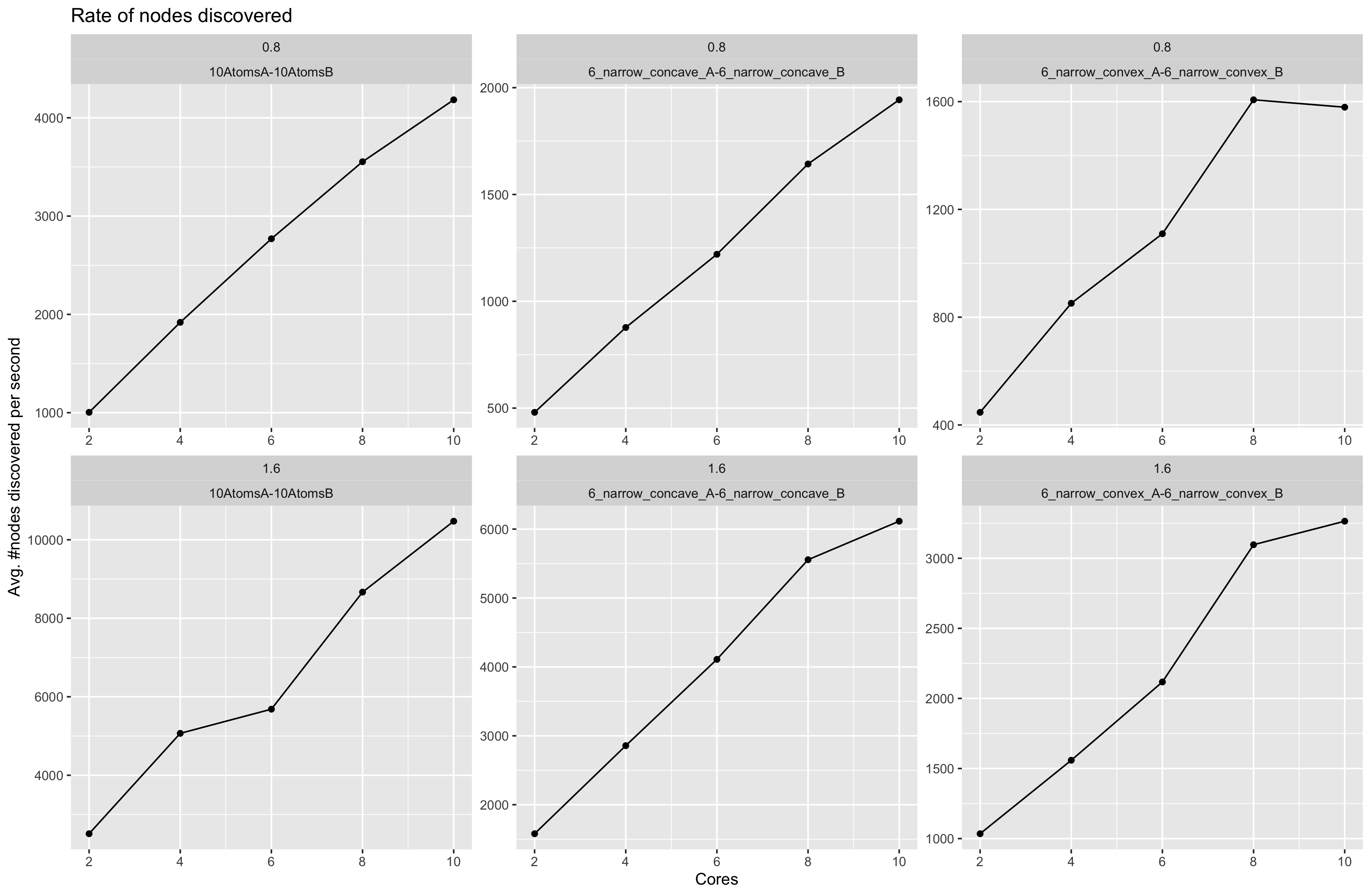}
\caption[Compute cores vs Node discovery rate]{Shows the number of compute cores on the
x-axis and the average number of macrostates or atlas nodes discovered per
second, while atlasing the two molecule mentioned at the top of every
sub-figure, for two different step sampling sizes. }
\label{fig:avgNodeDiscTime}
\end{figure}

Figure \ref{fig:avgNodeDiscTime} pltos the number of compute cores on the
x-axis and the average number of macrostates or atlas nodes discovered per
second, while atlasing the two molecule mentioned at the top of every
sub-figure, for two different step sampling sizes. As can be seen from the
figure, the number of nodes discovered per second increases with increase in
the number of cores.

\begin{figure}[htpb]
    \centering
    \includegraphics[scale=0.7]{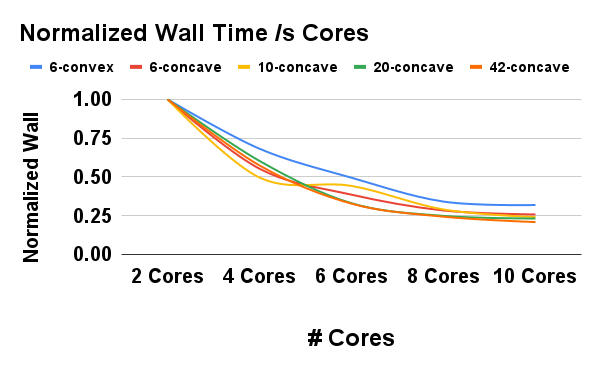}
    \caption{Plot of normalized wall time for atlas sampling on the y-axis, and the number of compute cores used on the x-axis}
    \label{fig:normalized_wall_time}
\end{figure}

In Figure \ref{fig:normalized_wall_time} we plot the normalized wall time 
for atlas sampling on the y-axis, and the number of compute cores used on the 
x-axis, for 6 different input molecules. The plot shows that the speedup achieved 
through parallelism is better for bigger molecules. This is expected, as the time 
for sampling each node is dependent on the square of the number of atoms (see 
complexity analysis in \cite{PrabhuEtAl2020}). Since this subroutine that depends 
heavily on the size of the molecule is being run in parallel, we see the highest 
benefit of parallelism when the subroutine takes the longest.

\begin{figure}[htpb]
    \centering
    \includegraphics[scale=0.7]{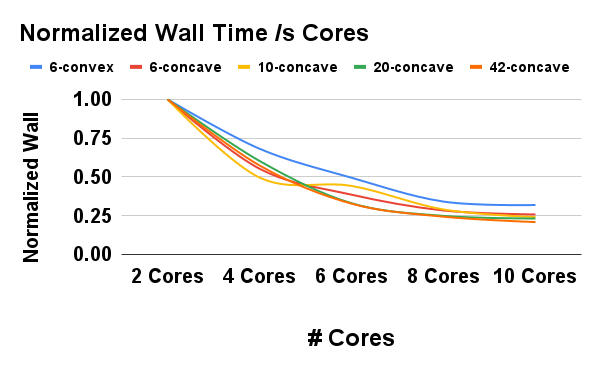}
    \caption{Plot of the number of compute cores used on the x-axis and the 
    normalized wall time for atlas sampling on the y-axis, with the 20 concave 
    molecule as the input and three different step sizes.}
    \label{fig:normalized_wall_time_steps}
\end{figure}

A similar behavior is seen in Figure \ref{fig:normalized_wall_time_steps} where 
we plot the normalized wall time for atlas sampling on the y-axis, and the number 
of compute cores used on the x-axis, for the same molecule but with different 
step sizes. The smaller the step size, the longer the sampling subroutine takes 
to sample each atlas node (see complexity analysis in \cite{PrabhuEtAl2020}). By 
parallelizing the sampling of atlas nodes, we see the highest benefit when the 
step size is the smallest.

\section{Discussion}
\label{sec:discussion}

\subsection{Profiling}
\label{sec:discussion:profiling}
\begin{figure}[htpb]
\centering
\includegraphics[width=\textwidth]{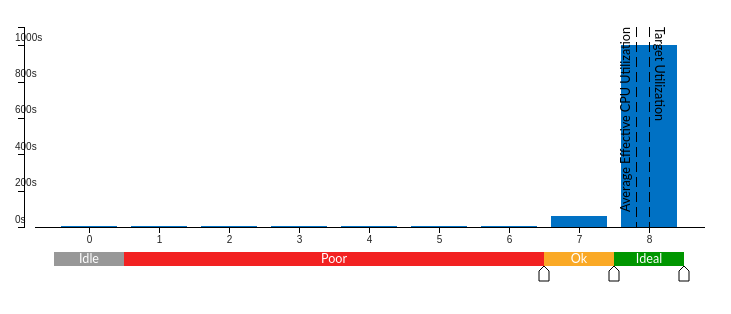}
\caption[Profile of the EASAL parallel code]{Profile of the EASAL parallel code. The x-axis is the number of cores
and the y-axis is the time spent on each core. The average effective CPU
utilization line is very close to the target utilization line, indicating that
the implementation is close to the theoretical limit of CPU utilization.}
\label{fig:profile}
\end{figure}

The code was profiled using Intel(R) Vtune (TM) parallel studio profiler.
As can be seen from Figure \ref{fig:profile}, our implementation of the 
parallel EASAL algorithm, shows close to the theoretical limit of CPU utilization
with 8 cores. However, this utilization reduces when more cores are added, plateauing
around a minimum imposed by the inherent level of parallelism in the problem. 

\subsection{Design Decisions}
\label{sec:designDecisions}
This section describes some of the non-intuitive design decisions taken to increase
the performance of the parallel algorithm.

\subsubsection{Single \aba}
\label{sec:designDecisions:numABA}
We use a single \aba\ in our implementation to keep track of the central copy
of the atlas and avoid repeat sampling macrostates. This is potentially a
performance bottleneck as all newly discovered regions have to be first
processed by the \aba\ before being sampled. This bottleneck can potentially
be alleviated by having multiple \aba s. However, there are some key
challenges that prevent this. With multiple \aba s, if we have a single copy of
the atlas data structure, they will need to share the data, requiring locks,
thus neutralizing the advantages of the asynchronous actor model. Maintaining
multiple copies of the atlas data structure will require distributed consensus
to keep the data structure updated across all the different \aba s. The third
optin is to partition the atlas data structure across the different \aba s,
which would still require a quick hashing to determine which \aba\ a newly
discovered atlas node should be sent to. Unfortunately, there is no easy way
to hash active constraint regions this way.

\subsubsection{Single \wa}
\label{sec:designDecisions:numWA}
We use a single \wa\ to serialize the writes to the disk.
Most hard-drives only support single channel input, which means that only
one thread can write to the disk at a time. If multiple writes are initiated
they need to be multiplexed, which substantially slows down the performance
of the disk in terms of write speeds. Even modern PCIe dual channel drives 
only support 2 channels and cannot handle tens of threads writing to them
without suffering a degradation in their write speeds.

\subsubsection{Limit on the Number of \sa s}
\label{sec:designDecisions:numSA}
Our experiments show that we get the best performance when we have the same
number of actors as the number of logical threads available. Also,
having more actors be active at the same time also increases the memory
required by the program.

\bibliographystyle{plain}
\bibliography{references}

\begin{thebibliography}{1}

\bibitem{chs-rapc-16}
Dominik Charousset, Raphael Hiesgen, and Thomas~C. Schmidt.
\newblock {Revisiting Actor Programming in C++}.
\newblock {\em Computer Languages, Systems \& Structures}, 45:105--131, April
  2016.

\bibitem{cshw-nassp-13}
Dominik Charousset, Thomas~C. Schmidt, Raphael Hiesgen, and Matthias
  W{\"a}hlisch.
\newblock {Native Actors -- A Scalable Software Platform for Distributed,
  Heterogeneous Environments}.
\newblock In {\em Proc. of the 4rd ACM SIGPLAN Conference on Systems,
  Programming, and Applications (SPLASH '13), Workshop AGERE!}, pages 87--96,
  New York, NY, USA, Oct. 2013. ACM.

\bibitem{ota2006multi}
Jun Ota.
\newblock Multi-agent robot systems as distributed autonomous systems.
\newblock {\em Advanced engineering informatics}, 20(1):59--70, 2006.

\bibitem{easalSoftware}
Aysegul Ozkan, Rahul Prabhu, Troy Baker, James Pence, and Meera Sitharam.
\newblock Efficient atlasing and search of assembly landscapes, 2021.
\newblock EASAL software. Available Online:
  https://bitbucket.org/geoplexity/easal.

\bibitem{easalVideo}
Rahul Prabhu, Troy Baker, and Meera Sitharam.
\newblock Video illustrating the opensource software {EASAL}, 2016.

\bibitem{easalUserGuide}
Rahul Prabhu and Meera Sitharam.
\newblock {EASAL} software user guide., 2016.

\bibitem{PrabhuEtAl2020}
Rahul Prabhu, Meera Sitharam, Aysegul Ozkan, and Ruijin Wu.
\newblock Atlasing of assembly landscapes using distance geometry and graph
  rigidity.
\newblock {\em Journal of Chemical Information and Modeling, to appear}, 2020.

\end{thebibliography}
\end{document}